\documentclass[amsmath,amssymb,
showpacs,twocolumn, nobibnotes,aps,prd]{revtex4-1}

\usepackage{graphicx,amsmath,amssymb}
\usepackage{mathrsfs,eufrak}
\usepackage{dcolumn,color}
\usepackage{bm}

\begin{document}


\title{Fully differential cross sections for $ Li^{2+}$-impact ionization of $Li(2s)$ and $Li(2p)$}


\author{Omid Ghorbani$^1$}
\author{Ebrahim Ghanbari-Adivi$^1$}\email{ghanbari@phys.ui.ac.ir}
\author{Marcelo Fabian Ciappina$^2$}

\affiliation{$^1$ Department of Physics, Faculty of Sciences,
University
of Isfahan, Isfahan 81746-73441, Iran\\
$^2$ Institute of Physics of the ASCR, ELI-Beamlines, Na Slovance 2,
182 21 Prague, Czech Republic}

\begin{abstract}
A semiclassical impact parameter version of the continuum distorted
wave-Eikonal initial state theory is developed to study the
differential ionization of $Li$ atoms in collisions with $Li^{2+}$
ions. Both post and prior forms of the transition amplitude are
considered. The fully differential cross sections are calculated for
the lithium targets in their ground and their first excited states
and for the projectile ions at 16~MeV impact energy. The role of the
internuclear interaction as well as the significance of the
post-prior discrepancy in the ejected electron spectra are
investigated. The obtained results for ejection of the electron into
the azimuthal plane are compared with the recent measurements and
with their corresponding values obtained using a fully quantum
mechanical version of the theory. In most of the cases, the
consistency of the present approach with the experimental and the
quantum theoretical data is reasonable. However, for $2p$-state
ionization, in the cases where no experimental data exist, there is
a considerable difference between the two theoretical approaches.
This difference is questionable and further experiments are needed
to judge which theory makes a more accurate description of the
collision dynamics.
\end{abstract}
\keywords{CDW-EIS. 3DW-EIS. Ionization. Binary peak. Fully
differential cross sections}
\pacs{34.80.Dp}
\maketitle
\section{Introduction~\label{Sec01}}
The study of the electronic reactions due to the collision of
electrons and heavy ions with atomic and molecular targets is one of
the interesting topics in atomic and molecular physics that has
attracted considerable attention for many years. Apart from the
challenges that appear in both experimental and theoretical
investigations of such phenomena, there exist some other reasons
that highlight the importance of these processes. For example, these
reactions play a predominant role in applied areas such as the
design of fusion reactors, the study of heavy-particle radiation
damage in human tissues and the application of hadron therapy. Also,
in many scientific areas such as astrophysics and plasma physics, a
detailed knowledge of the mechanisms of these processes is usually
required~\cite{Stolterfoht,Drake}. However, one of the main aspects
of these phenomena is the fact that they provide an interesting
probe of the few-body problem~(FBP) in physics~\cite{Schulz1}. The
Coulomb interaction which is responsible for these reactions is
exactly known, while the Schr\"{o}dinger equation governing the
dynamics of such systems is not analytically solvable for more than
two mutually interacting particles. Consequently, the validity of
the various approximate theories which are developed to shed light
on questions relevant to dynamics of the few-body Coulomb systems
should be assessed by comparing their predictions with precise
experimental data.\par
When an energetic bare ion impacts an atomic or a molecular target,
there is a definite probability for occurrence of each of the basic
processes such as electron capture and excitation or ionization of
the target. Among these various processes, ionization in particular
has received a great part of attention, because not only it is the
main mechanism leading to the energy loss of the swift ions in
matter but also, it provides an ideal testing ground for the Coulomb
FBP~\cite{Schulz2,Rescigno}. This is especially true  for the single
ionization of the target for which there are three unbound particles
in the final state~\cite{Madison1}. More challenge to theoretical
models for proving their accuracy comes from the measured fully
differential cross sections~(FDCSs), because the more differential
cross sections, the more information that can be derived about the
collision mechanism. In the case of electron impact, the
kinematically complete experiments have provided a remarkable amount
of the measured FDCSs to investigate the few-body aspects of the
$(e,2e)$ processes~\cite{Ehrhardt,Lahmann,Roder1,Dorn,Haynes}. In
such experiments, the complete momentum vectors of all the collision
fragments are determined. For the case of ion impact, since both the
scattering angles and the changes in the magnitude of the projectile
momentum  are very small, these experiments are much more
challenging. It took a long time after the corresponding experiment
for electron impact which the difficulties associated with the large
projectile mass were circumvented by measuring the momentum vectors
of the recoiling target ion and the ejected electron, using the
novel method of the cold-target recoil-ion momentum spectroscopy
(COLTRIMS)~(also known as a reaction
microscope~(ReMi))~\cite{Schulz1,Ullrich,Dorner}. The scattered
projectile momentum can then be deduced from the kinematic
conservation laws. COLTRIMS is limited to gaseous targets of
relatively small mass number. In order to overcome this limitation,
the supersonic gas jet applied to produce the very cold target beam
in a conventional reaction microscope was replaced by a
magneto-optical trap~(MOT) to innovate the MOTReMi
method~\cite{Fischer1,Hubele1}. This method improved the achievable
recoil-ion momentum resolution considerably and extended the targets
to heavy  atoms or molecules which can be optically pumped.\par
Development of the above mentioned experimental techniques has
renewed the interest in the theoretical study of the ion-impact
ionization of the atomic and molecular targets. A number of the
theoretical studies engaging these reactions have been done in
framework of the time dependent close coupling~(TDCC)~\cite{Colgan},
classical trajectory Monte Carlo~(CTMC)~\cite{Olson,Sarkadi},
continuum distorted wave-Eikonal initial
state~(CDW-EIS)~\cite{Ciappina1,Gulyas1}, coupled
pseudostate~(CP)~\cite{Walters1,Walters2} and three-body distorted
wave-Eikonal initial state (3DW-EIS)~\cite{Ghanbari1} theories.\par
Very recently, the measured FDCSs for 16~MeV~$Li^{2+}$ single
ionization of the $2s$ ground and the $2p$ excited states of lithium
has been reported and compared with the 3DW-EIS
calculations~\cite{Ghanbari2}. The study of this reaction is of
interest for several reasons. First, the radial nodal structure of
the $2s$ and $2p$ wavefunctions as well as the angular distribution
of the $2p$ wavefunction of the valence electron may lead to some
structures in the FDCS. Second, in the lithium targets, the inner
shell is both spatially and energetically far away from the valence
shell which includes a single valance electron. So, it is expected
that the electron-electron correlation plays a marginal role in the
collision dynamics. Additionally, for this system  the perturbation
parameter~(projectile charge to speed ratio) is much larger than
that for a typical~$(e,2e)$ reaction. This in combination with the
more complex structure of the wavefunctions can lead to new features
not seen previously.  The reported data are for ejection into the
azimuthal plane and the comparisons showed that the $2s$ and $2p$
cross sections exhibit different behaviors in some features. For
example, for the $2p$ case, a double binary peak structure was
theoretically predicted for some values of the projectile momentum
transfer and the ejected electron energy, while such a structure is
absent for some other values of these quantities. Using the 3DW-EIS
method, it was shown that the double peak structure is associated to
the kinematics of the reaction and to the angular part of the
wavefunction. Although, the good overall agreement was found between
the 3DW-EIS calculations and the experimental data, there exist some
discrepancies which motivate further investigation of the
process.\par
In this contribution, we apply the semiclassical CDW-EIS
method~\cite{Crothers1,Crothers2} to calculate the fully
differential cross sections for single ionization of $Li$ in
collision with 16~MeV~$Li^{2+}$ projectile ions. The followed
approach in fact provides an approximate solution to the
Schr\"{o}dinger equation governing the dynamics of the specified
collisional breakup process which in turn is approximated to a
three-body reaction with effective Coulomb interactions. The
solution satisfies both the initial and final correct boundary
conditions. In this treatment a classical straight-line trajectory
is considered for the projectile, while the states of the involving
particles are described quantum mechanically.  For an impact energy
of 16~MeV, assuming a classical straight-line trajectory for the
projectile is not too far from reality. Also, as is mentioned above,
the single optical active electron in $Li$ atoms is spatially far
away from the spherically symmetric full inner shell closest to the
nucleus. So, approximating the core interaction with the active
electron and the projectile ion as the Coulomb interactions~(with an
effective charge) is not too rough. Indeed, in the present study, we
use a linear combination of the hydrogenic wave functions,
Roothaan-Hartree-Fock~(RHF) wave functions~\cite{Clementi}, to make
an analytical fit to the numerical Hartree-Fock~(HF) wave functions.
It is well known that the three-dimensional wave functions can be
approximated as an expansion in terms of the hydrogenic wave
functions as a complete set. These assumptions, may make the
calculations easier, but it does not mean that such a complicated
problem is reduced to a simple hydrogenic problem. The obtained
results are discussed in comparison with experimental data and with
the calculations of the full quantum mechanical version  of the
theory.  Comparison shows that the mentioned assumptions give a
reasonable estimate of the FDCSs.\par
 Over the years, it has been shown that CDW-EIS theory
is one of the most successful approaches to explain the dynamics of
ion-atom collisional processes in the perturbative regime. We
considered a number of assumptions to perform the CDW-EIS
calculations easier and faster to do, and the reported results show
that these assumptions do not significantly affect the accuracy of
the calculations. Our further study on $Li^{2+}$-$Li$ system using
3DW-EIS shows that if we use an analytical fit to the numerical HF
wavefunction and employ a Coulomb potential to approximate the
interaction with the passive electron, the obtained results will not
change significantly in shape. These facts motivated us to use an
analytical RHF wave function for the initial electronic state and
use the Coulomb potentials to describe the interaction of the target
core with the projectile and the active electron. This
simplification reduces the computing time considerably. For example,
for the $2p$-state ionization and for each specified values of the
ejected-electron's energy and the projectile momentum transfer, the
3DW-EIS calculations occupy 60 CPUs for nearly 45 hours, while with
our present CDW-EIS approach the same results are obtained using
only one CPU through several minutes. \par
The plan of the paper is as follows. Section~\ref{Sec02} presents
the impact parameter version of the CDW-EIS model to investigate a
typical ion-impact ionization process. Section~\ref{Sec03} is
devoted to the results and the relevant discussions, and
section~\ref{Sec04} includes the summary and concluding remarks.
Atomic units $(e=m_e=\hbar=1)$ are used unless otherwise stated.\par
\section{Theory\label{Sec02}}
The continuum distorted wave-Eikonal initial state (CDW-EIS)
approach was firstly introduced by Crothers and his
coworkers~\cite{Crothers1,Crothers2} and applied as a successful
method to treat a wide variety of collision
systems~\cite{Ciappina1,Gulyas1,Salin,Fainstein,Busnengo,Rodriguez,Gulyas2,Fiol,Voitkiv1,Voitkiv2,Abufager,Voitkiv3,Monti,Belkic,Ciappina2,Voitkiv4},
so only a brief outline will be given here.\par
A number of simplifying assumptions is convenient in theoretical
investigation of a typical ion-atom collision system. The
assumptions which are used in the present model are as follows. (a)
We use the frozen-core approximation in which one of the electrons
of the target is considered as the active electron and the others
are considered frozen in their initial states as the passive ones.
The influence of the passive electrons is considered through the
effective screening potential in the calculations. (b) In the
semi-classical CDW-EIS model, the impact-parameter approximation is
used. In this approximation, a straight-line trajectory is
considered to describe the relative motion of the nuclei.
Consequently, the inter-nuclear position vector is parameterized by
the impact parameter $\mbox{\boldmath$\rho$}$ and the constant
relative velocity ${\bf v}$ as ${\bf R}(t)=\mbox{\boldmath$\rho$} +
{\bf v} t$ with $\mbox{\boldmath$\rho$}\cdot{\bf v}=0$. Hence, the
closest distance between the heavy particles occurs at the time
$t=0$. (c) The RHF~wavefunctions are employed to describe the state
of the bound subsystem in the entrance channel~\cite{Clementi}. (d)
In the final channel, the potential interaction between the ionized
electron and the residual target ion is approximated by an effective
Coulomb potential. (e) the final continuum-state of the ionized
electron in the field of the residual target ion is considered as a
Coulomb wave.\par
By these assumptions and working in the framework of the
independent-electron frozen-core model, the collision system can be
reduced to a three-body system including a bare projectile ion $P$,
with charge of $Z_P$ and mass of $M_P$, the active electron $e$, and
the target ion $T$ with an effective charge of $Z_{T}$ and mass of
$M_T$. It is assumed that $P$ impinges on $(T+e)$ with the initial
wave vector ${\bf K}_i$. Over the course of the collision, $P$ is
scattered with the final momentum ${\bf K}_f$ and $e$ is ejected
with wave vector ${\bf k}_e$. The full electronic Hamiltonian of
these three-body system reads~\cite{Fainstein}
\begin{equation}\label{EQ01}
H_{el} = H_0 + V_P(|{\bf R}-{\bf r}|)+ V_S (R),
\end{equation}
in which ${\bf r}$ is the position vector of the active electron
with respect to $T$, $H_0$ is the Hamiltonian for the free target
atom and $V_P(|{\bf R}-{\bf r}|)$ and $V_S (R)$ denote the $P$-$e$
and $P$-$T$ interactions. The effective nucleus charge is considered
as $Z_{eff}=n_i\sqrt{-2\varepsilon_i} $ in which $n_i$ is the
principal quantum number of the electron orbit and $\varepsilon_i$
is its binding energy. Whit this assumption $V_S (R)$ can be
approximated by $V_S(R) = Z_{eff}Z_P/R$. Within the present
framework, this fact that $V_S(R)$ depends only on the inter-nuclear
distance gives rise to a phase factor which corresponds to the
important effect of the nuclear-nuclear interaction~(NN~interaction)
on FDCSs~\cite{Ciappina1}.\par
In this step, we withdraw the NN~interaction from the total
electronic Hamiltonian. Hence, $H_{el}$ can be rearranged
as~\cite{Fainstein}
\begin{equation}\label{EQ02}
\begin{split}
H_{el} & = H_i \,+ U_i \, + W_i \, = H_i^d + W_i,\cr H_{el} & = H_f
+ U_f + W_f = H_f^d + W_f,
\end{split}
\end{equation}
where $H_i~(H_f)$ is the Hamiltonian for the subsystem $(e+T)$ in
the entrance~(exit) channel. The distortion potentials $U_i$ and
$U_f$ are defined in such a manner that the Schr\"{o}dinger
equations corresponding to the distorted Hamiltonians, $H_i^d=H_i
\,+ U_i $ and $H_f^d = H_f + U_f$, are exactly solvable.
Consider $\chi_i^+({\bf r},t) $ and $\chi_f^-({\bf r},t)$ as the
exact solutions of the Schr\"{o}dinger equations,
\begin{equation}\label{EQ03}
\big(H_i^d-i{\partial\over\partial t}\big)\chi_i^+({\bf r},t)=0,\
{\rm and}\ \big(H_f^d-i{\partial\over\partial t}\big)\chi_f^-({\bf
r},t)=0,
\end{equation}
that satisfy the outgoing and incoming boundary conditions,
respectively. Having these solutions, the perturbation potentials
$W_i$ and $W_f$ are obtained using equations
\begin{equation}\label{EQ04}
\begin{split}
\big( H_i - E_i \big ) \chi_i^+({\bf r},t) & = W_i \chi_i^+({\bf
r},t),\cr \big( H_f - E_f )\chi_f^-({\bf r},t)& =W_f \chi_f^-({\bf
r},t),
\end{split}
\end{equation}
where $E_i~(E_f)$ is the total  initial~(final) energy of the system
in the center-of-mass~(c.m.) frame. These potentials are weaker than
those appearing in the Born series, so it is expected that the
distorted wave series corresponding to these perturbation potentials
converge faster than those appear in Born approximation. Also,
assume that $\psi_i^+({\bf r},t)$ and $\psi_f^-({\bf r},t)$ are
respectively the incoming and outgoing solutions of the
Schr\"{o}dinger equation
\begin{equation}\label{EQ05}
\big(H_{el}-i{\partial\over\partial t}\big)\psi({\bf r},t)=0,
\end{equation}
with the exact asymptotic behaviors~\cite{Fainstein,Belkic}. The
above introduced wavefunctions satisfy the conditions
\begin{equation}\label{EQ06}
\mathop{\lim}\limits_{t\to-\infty} \langle \psi_f^-|\chi_i^+\rangle
= 0,\qquad \mathop{\lim}\limits_{t\to+\infty}^{}\langle\chi_f^-|
\psi_i^+\rangle = 0,
\end{equation}
which imply that the transition cannot be produced by the action of
the distortions.\par
In the present version of CDW-EIS,  $\chi_i^+({\bf r},t)$ and
$\chi_f^-({\bf r},t)$ are approximated as~\cite{Fainstein}
\begin{equation}\label{EQ07}
\begin{split}
\chi_i^+({\bf r},t) & = \Phi_i^+ ({\bf r},t){\cal L}_i^ {+EIS} ({\bf
r}), \cr \chi_f^-({\bf r},t) & = \Phi_f^- ({\bf r},t){\cal L}_f^
{-CDW} ({\bf r}),\end{split}
\end{equation}
where $\Phi_i^+ ({\bf r},t)$ and $\Phi_f^- ({\bf r},t)$ are the
incoming and outgoing wavefunctions in the first Born approximation
\begin{equation}\label{EQ08}
\begin{split}
\Phi_i^+({\bf r},t) & = \psi_i({\bf r}_T) e^{-i({1\over 2}{\bf
v}\cdot{\bf r}+{1\over 8}{\rm v}^2t+\epsilon_i t)} \cr \Phi_f^-
({\bf r},t) & =\psi_{C}^-({\bf r}_T)e^{-i({1\over 2}{\bf v}\cdot{\bf
r}+{1\over 8}{\rm v}^2t+E_f t)},
\end{split}
\end{equation}
where $\epsilon_i$ is the initial binding energy of the electron,
$\psi_i({\bf r}_T)$ is the wavefunction describing its initial bound
state. $\psi_C^-({\bf r}_T)$ is the Coulomb distorted wavefunction
describing the motion of the ejected electron in the field of the
residual target ion with an explicit form of
\begin{equation}\label{EQ09}
\begin{split}
\psi_C^- ({\bf r}_T) & = (2\pi)^{-3/2}N(\nu_T)e^{i{\bf k}_e\cdot
{\bf r}_T}\cr &\times {_1F_1}(i\nu_T;1;-i k_e r_T -i {\bf k}_e\cdot
{\bf r}_T).
\end{split}
\end{equation}
${\bf k}_e$ is the momentum vector of the ejected electron with
respect to the target ion, ${_1F_1}(a;b;z)$ is the confluent
hypergeometric function and the usual normalization factor
$N(\nu_T)$ is given by $N(\nu_T)=\Gamma(1-i\nu_T) e^{-\pi\nu_T/ 2}$
in which $\Gamma(z)$ is Gamma function and $\nu_T=-Z_T/k_e$ is a
Sommerfeld parameter.\par
The distortion factors of ${\cal L}_i^ {+EIS} ({\bf r})$ and ${\cal
L}_f^ {-CDW} ({\bf r})$ are chosen so that the distorted waves,
$\chi_i^+({\bf r},t)$ and $\chi_f^-({\bf r},t)$, satisfy the correct
boundary conditions. The eikonal initial state distortion is a
logarithmic distortion phase factor due to the Coulomb long-range
remainder of the electron-projectile perturbation potential in the
initial channel. From Eqs.~(\ref{EQ03}) and~(\ref{EQ07}), this
distortion factor can be derived as ${\cal L}_i^{+EIS} ({\bf r})=
e^{-i \nu_i \ln ({\rm v} r_P+ {\bf v}\cdot{\bf r}_P)},$ where
$\nu_i=Z_P/{\rm v}$. Similarly, it can be shown that the Coulomb
distortion factor representing the post-collision interaction~(PCI)
between the projectile ion and the active electron is a proper
expression for ${\cal L}_f^ {-CDW} ({\bf r})$. So, we have
\begin{equation}\label{EQ10}
{\cal L}_f^{-CDW} ({\bf r})=N(\nu_P) {_1F_1}\big(i\nu_P;1;-i
k_Pr_P-i {\bf k}_P\cdot{\bf r}_P),
\end{equation}
where $\nu_P=-Z_P/k_P,$ and ${\bf k}_P={\bf k}_e-{\bf v}$ is the
momentum of the ejected electron with respect to the projectile
nucleus.\par
The perturbation potential corresponding to the above forms of the
distorted waves and the distortion factors are given
by~\cite{Fainstein}
\begin{equation}\label{EQ11}
W_i  = {1\over 2} \nabla_{r_P}^2
+\nabla_{r_P}\cdot\nabla_{r_T},\quad {\rm and}\quad W_f =
\nabla_{r_P}\cdot\nabla_{r_T},
\end{equation}
which are consistent with Eq.~(\ref{EQ04}).\par
The post and prior versions of the ionization probability amplitude,
described from a reference frame fixed on the target nucleus, can be
respectively written as~\cite{Fainstein}
\begin{equation}\label{EQ12}
\begin{split} a_{fi}^{+}(\mbox{\boldmath$\rho$} ) & =  - i \int_{-\infty}^{+\infty}  dt \langle
\chi_f^- |W_f^\dag | \psi_i^+ \rangle\cr   & \simeq - i
\int_{-\infty}^{+\infty} dt \langle \chi_f^- |W_f^\dag | \chi_i^+
\rangle,\cr a_{fi}^{-}(\mbox{\boldmath$\rho$} ) & = - i
\int_{-\infty}^{+\infty}  dt \langle \psi_f^- |W_i | \chi_i^+
\rangle \cr & \simeq - i \int_{-\infty}^{+\infty}  dt \langle
\chi_f^- |W_i | \chi_i^+ \rangle.
\end{split}
\end{equation}
NN~interaction is not included in these versions of the transition
amplitude. Also, in an exact treatment, these forms of the amplitude
are mathematically equivalent, while in an approximate treatment a
post-prior discrepancy may be expected.\par
Now, we are ready to take into account the influence of the
NN~interaction on the probability amplitude.  Using the procedure
outlined in Refs.~\cite{Crothers2,Dewangan}, this influence can be
considered by multiplying the above amplitudes by a phase factor due
to the pure Coulomb interaction between $P$ and $T$. Accordingly,
the transition amplitude including the NN-interaction influence
reads
\begin{equation}\label{EQ13}
{\cal A}_{fi}^\pm(\mbox{\boldmath$\rho$}) = i (\rho {\rm
v})^{2i\nu}a_{fi}^\pm (\mbox{\boldmath$\rho$}),
\end{equation}
where  $\nu = Z_P Z_T/{\rm v}$ is a Sommerfeld parameter, and $Z_T$
is the asymptotic charge of the target ion $T$.\par
Using the two-dimensional Fourier transforms, the corresponding
probability amplitudes in the momentum-space can be obtained
as~\cite{Ciappina1,Dewangan}
\begin{equation}\label{EQ14}
\begin{split}
t_{fi}^\pm(\mbox{\boldmath$\eta$})  = {1\over 2\pi}  \int d
\mbox{\boldmath$\rho$}
e^{i\mbox{\boldmath$\eta$}\cdot\mbox{\boldmath$\rho$}} a_{fi}^\pm
(\mbox{\boldmath$\rho$}), \cr {\cal
T}_{fi}^\pm(\mbox{\boldmath$\eta$}) = {1\over 2\pi} \int d
\mbox{\boldmath$\rho$}
e^{i\mbox{\boldmath$\eta$}\cdot\mbox{\boldmath$\rho$}} {\cal
A}_{fi}^\pm (\mbox{\boldmath$\rho$}),
\end{split}
\end{equation}
in which $\mbox{\boldmath$\eta$}$ is the transverse momentum
transfer. Inserting the inverse Fourier transform of
$t_{fi}^\pm(\mbox{\boldmath$\eta$})$ into~(\ref{EQ13})  and getting
the Fourier transform of the result leads to
\begin{equation}\label{EQ15}
{\cal T}_{fi}^\pm(\mbox{\boldmath$\eta$}) = {i{\rm v}^{2i\nu} \over
(2\pi)^2} \int d \mbox{\boldmath$\eta$}'
t_{fi}^\pm(\mbox{\boldmath$\eta$}')\int d \mbox{\boldmath$\rho$}
\rho^{2i\nu}
e^{i(\mbox{\boldmath$\eta$}-\mbox{\boldmath$\eta$}')\cdot\mbox{\boldmath$\rho$}}.
\end{equation}
The integral over $\mbox{\boldmath$\rho$}$ can be analytically
performed to obtain~\cite{Ciappina1}
\begin{equation}\label{EQ16}
{\cal T}_{fi}^\pm(\mbox{\boldmath$\eta$}) = {\nu\over 2 (2\pi)^3}
\int d \mbox{\boldmath$\eta$}' t_{fi}^\pm(\mbox{\boldmath$\eta$}')
\big|\mbox{\boldmath$\eta$}-\mbox{\boldmath$\eta$}'\big|^{-2(1+i\nu)},
\end{equation}
in which the overall phase factor of $i({\rm v}^2/2\pi)^{i\nu}$ is
dropped since it affects neither the probability nor the cross
section of the reaction. The integral over $\mbox{\boldmath$\eta$}'$
can be performed numerically to obtain the triple differential cross
section for ionization of the active electron.\par
The fully differential cross section~(FDCS) for the considered
breakup process is given as~\cite{Voitkiv5}
\begin{equation}\label{EQ17} {d^5\sigma\over d^2\mbox{\boldmath$\eta$}
d^3{\bf k}_e} = {1\over 4\pi^2 \rm v} \big| {\cal
T}_{fi}(\mbox{\boldmath$\eta$})\big|^2
\end{equation}
in which ${\cal T}_{fi}(\mbox{\boldmath$\eta$})$ is the
corresponding post or prior transition matrix element. Since
$d^2\mbox{\boldmath$\eta$} d^3{\bf k}_e$ can be expressed in terms
of the triple differential $d E_k d \Omega_e \Omega_P$, this cross
section is more commonly referred to as triply differential cross
sections (TDCS) for ejection of the electron with energy between
$E_e$ and $E_e+dE_e$ into solid angle $d\Omega_e$ in direction of
${\bf k}_e$ and scattering of the projectile into solid angle
$d\Omega_P$ in direction of ${\bf K}_f$~\cite{Fischer}.\par
\begin{figure*}[t]
\begin{center}
\includegraphics[scale=0.97]{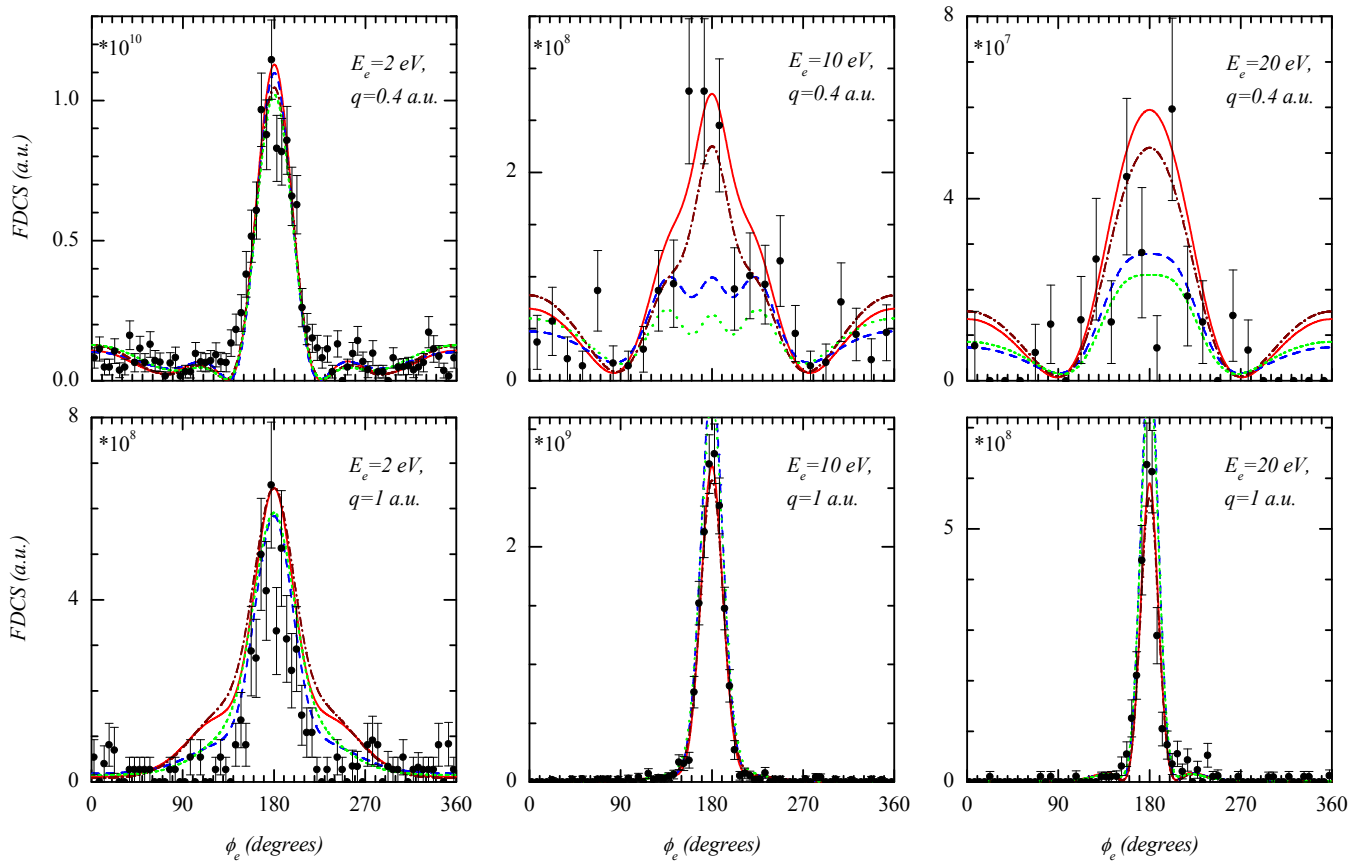}
\caption{(Color online) CDW-EIS FDCSs for ionization of the ground
state lithium atoms are compared with their experimental
values~\cite{Ghanbari1}. The solid~(red) curves are for  prior
CDW-EIS with NN interaction, the dashed~(blue) lines are the same
but for post CDW-EIS. The dash-dotted~(brown) curves are for prior
calculations without NN interaction and the dotted~(green) lines are
the same for post calculations. Closed symbols are for
experiments~\cite{Ghanbari2}.}\label{Fig01}
\end{center}
\end{figure*}
At the end, let us refer to the main differences between the present
distortion method and its fully quantum mechanical version which is
known as 3DW-EIS in the literature.  First, in the present model a
classical straight trajectory is assumed for the projectile while
such an assumption is not allowed in quantum mechanics. Second, we
have used analytical wavefunctions for the initial bound and the
final continuum stats of the active electron, while the numerical HF
wavefunctions are used for both in 3DW-EIS. Third, in 3DW-EIS
amplitude there exists an extra term coming from a proper
rearrangement of the terms included in the exact amplitude, and
taking into account the additional higher-order effects through the
final-state perturbation, while this term is absent in CDW-EIS.\par
\section{Results\label{Sec03}}
In this section, we use the above outlined method to study the
16~MeV $Li^{2+}$-impact differential single ionization of the
ground~($2s$) and the first excited~($2p$) states of $Li$ targets.
To this end, the FDCSs are calculated as a function of the ejected
electron's azimuthal angle for ejection into the plane perpendicular
to the incident projectile beam. The results are compared with the
experimental data as well as the results obtained using the fully
quantum mechanical version of the theory~(or 3DW-EIS theory), which
both have been recently reported in Ref.~\cite{Ghanbari2}. We choose
the origin of the coordinate system at the target nucleus $T$,  with
the $z$-axis along the incident beam. It is assumed that the
projectile is scattered in the $+xz$ half plane and the positive
direction of the $y$ axis is chosen so it makes a right-handed
coordinate system. The azimuthal and polar angles of the ejected
electron is measured as usual with respect to the positive
directions of the $x$ and $z$ axes, respectively.\par
 In addition to the ionization occurring in the
$Li^{2+}$-$Li$ collisional system, some other processes like
electron capture by $Li^{2+}$, excitation or ionization of the inner
shell electrons in the target atom and even the ionization of the
projectile ion may occur during the collision process. However,
since our considered experimental data are for the outer shell
ionization of the target, we neglected the other possible
reactions.\par
\begin{figure*}[t]
\begin{center}
\includegraphics[scale=0.97]{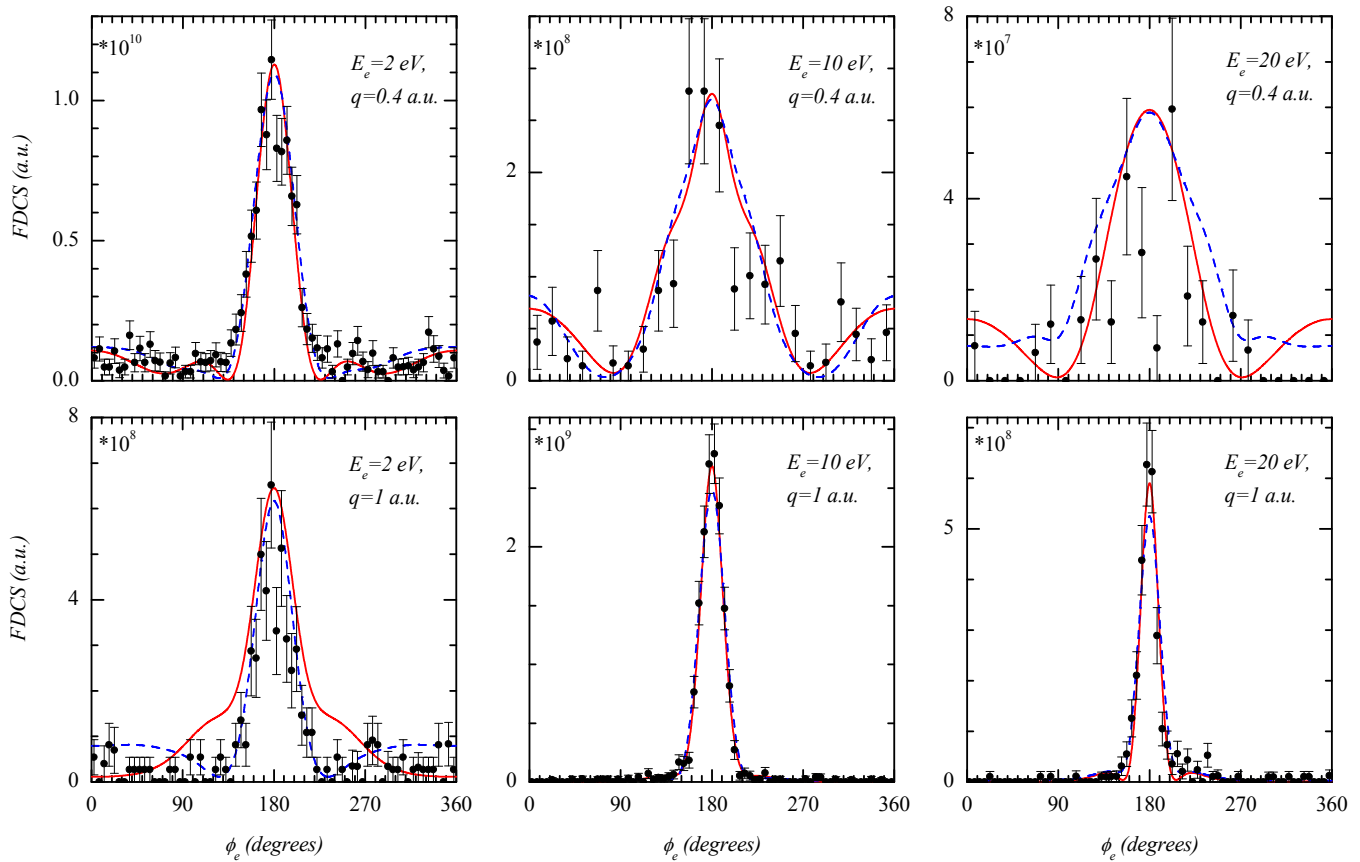}
\caption{(Color online) Comparison of the full prior CDW-EIS cross
sections with their corresponding values obtained using the 3DW-EIS
model and with the measured data. Solid~(red) curves are for
CDW-EIS, dashed~(blue) curves for 3DW-EIS and closed symbols for
experiments.}\label{Fig02}
\end{center}
\end{figure*}
For three different values of the ejected-electron's energy,
$E_e=$~2,~10, and 20~eV, and two different values of the projectile
momentum transfer, $q=|{\bf K}_i-{\bf K}_f|=$~0.4, and 1~$a.u.$, the
calculated post and prior FDCSs for single ionization of lithium
atoms in their ground states are compared with their experimental
values in figure~\ref{Fig01}. Both post and prior cross sections
with and without NN interaction are included in this figure. In
order to do a meaningful comparison, in each case the prior results
with the NN interaction are normalized to the experiment at the
binary peak, and the other results are multiplied by the same
normalization factor. Considering the figure, several points are
remarkable: as is kinematically expected, in accordance with the
experiments, the theory predicts a binary peak at
$\phi_e=180^\circ$. For cases ($E_e=2$~eV, $q=0.4~a.u.$),
($E_e=10$~eV, $q=1~a.u.$), and ($E_e=20$~eV, $q=1~a.u.$), the NN
interaction has no considerable influence on the cross sections and
the post-prior discrepancy is very small. However, for these cases,
maximum of the both of these effects occurs at the binary peak. For
($E_e=2$~eV, $q=1~a.u.$), the post-prior discrepancy is observed
both at  $\phi_e=180^\circ$ and at two mirror-symmetric angular
regions on both sides of the binary peak. In this case, the NN
interaction plays a minor role. For $E_e=$~10 and 20~eV cases with
$q=0.4~a.u.$, both the post-prior discrepancy and the NN-interaction
influence are quite obvious. This means that for higher emission
energies and lower momentum transfers, the influence of both the NN
interaction and the post-prior discrepancy on the cross sections is
considerable. This effect becomes much more important around the
binary peak for which a structure is seen in the results. For the
cases ($E_e=10$~eV, $q=1~a.u.$) and ($E_e=20$~eV, $q=1~a.u.$) around
the binary peak, the prior values are upper than the post results,
while the situation is reversed for the same energies but the lower
momentum transfers. It is interesting that the shape of the post and
prior results is nearly the same in all cases, except for
($E_e=10$~eV, $q=1~a.u.$) and ($E_e=20$~eV, $q=1~a.u.$). Although,
on the whole, good overall agreement is found between the
calculations and the experimental data in all cases, it seems the
prior calculations with considering the NN interaction gives a
better description of the collision dynamics. It is probable that
the results would be improved both in shape and in magnitude by
considering the procedure outlined in Refs.~\cite{Abufager}
and/or~\cite{Monti}.\par
\begin{figure*}[t]
\begin{center}
\includegraphics[scale=1]{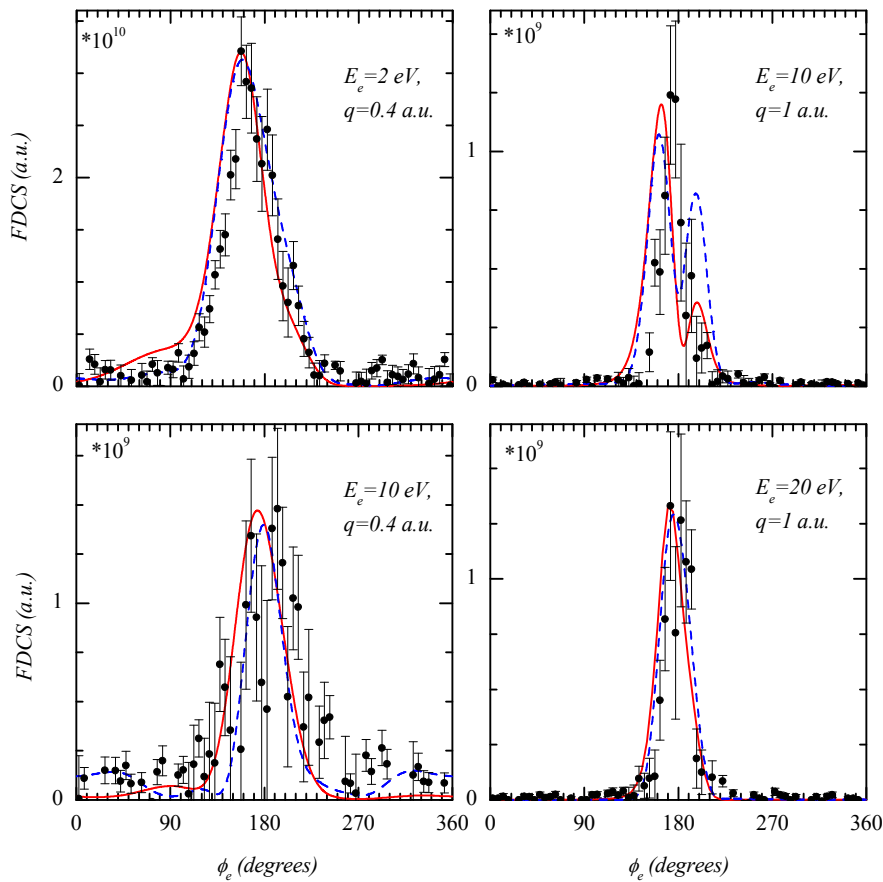}
\caption{(Color online) Same as figure~\ref{Fig02} but for
ionization of $Li(2p)$ targets.}\label{Fig03}
\end{center}
\end{figure*}
  It should be also remarked that in order to keep the
present model in a three-body framework, the projectile  is assumed
as a bare ion. Although the obtained results show that this
assumption is not too far from reality, the influence of the
projectile-electron on the cross sections can be examined using a
more complicated four-body model. For the cases that agreement of
the calculations with experiments is poor, such an approach probably
improves the results, but it is not followed here.  \par
FDCSs for ionization of $Li(2s)$ targets calculated using the prior
version of CDW-EIS including the NN interaction are compared with
their corresponding results obtained using the 3DW-EIS
model~\cite{Ghanbari2} and with their experimental
values~\cite{Ghanbari2} in figure~\ref{Fig02}. The results of both
theories are normalized to the experiments at the binary peak. As is
seen from this figure, it is interesting that in the cases for which
the influence of the NN-interaction and post-prior discrepancy on
the cross sections is insignificant, the theories completely
coincide with each other at all the azimuthal angles. In these
cases, both theories are also in good accordance with experiments.
The most difference between the semiclassical CDW-EIS and the full
quantum 3DW-EIS theories occurs for lower ejection energies but
higher momentum transfers such as ($E_e=2$~eV, $q=1~a.u.$) and for
higher energies but lower momentum transfers such as ($E_e=20$~eV,
$q=0.4~a.u.$). For these cases, it seems  that 3DW-EIS is in better
agreement with the measured data.\par
A similar comparison is made for single ionization of the $2p$
excited states of $Li$ targets in figure~\ref{Fig03}. Since the
targets were experimentally prepared~\cite{Ghanbari2} with relative
populations of $86\%$, $9\%$, and $5\%$ for the magnetic substates
with $m = -1$, $m=0$, and $m=-1$, respectively, the cross sections
are calculated using these experimental weights. If the sublevels
with $m=\pm 1$ were equally populated, a symmetric angular
distribution is expected for both the calculated and the measured
cross sections around $\phi_e=180^\circ$. But for this case, since
these two substates are not equally populated, this distribution is
asymmetric. This behavior which is known as magnetic dichroism~(or
orientational dichroism) in the angular distribution of  the
electron emission is exhibited in figure~\ref{Fig03}.\par
Like the case of the $2s$ state,  the post-prior discrepancy is also
seen for ionization of the $2p$-excited state, but it is much
smaller than that was found in figure~\ref{Fig01} for the ground
$2s$ state. Also, for all the specified cases represented in
figure~\ref{Fig02}, the influence of the NN~interaction on the cross
sections is unimportant. Consequently, the post and prior CDW-EIS
cross sections with and without NN interaction are nearly the same.
For this fact, only the prior CDW-EIS including the NN interaction
(the full prior CDW-EIS) results are illustrated in the figure.\par
\begin{figure*}[t]
\begin{center}
\includegraphics[scale=1]{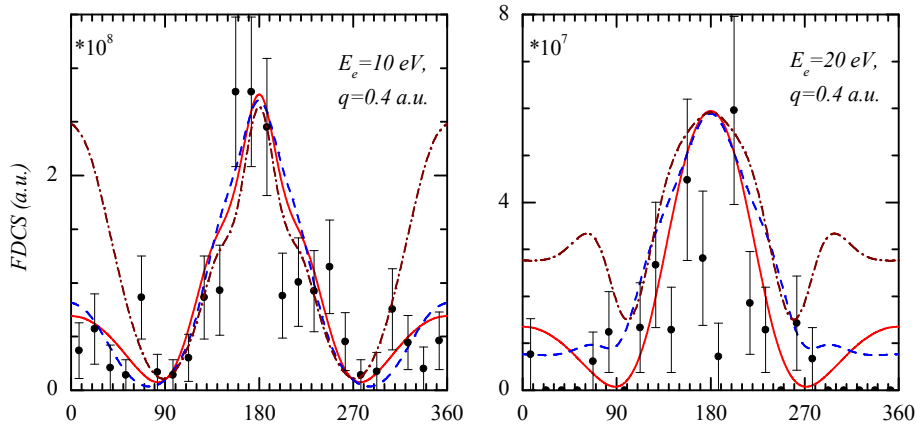}
 \caption{(Color online) Comparison of experimental and
theoretical 3DW-EIS, CDW-EIS and FBA FDCS results for 16~MeV
$Li^{2+}$ ionization of the $2s$ shell of $Li$ in the azimuthal
plane as a function of the azimuthal angle. Symbols are for
experiments, solid~(red) curves for the prior CDW-EIS, dashed~(blue)
curves for 3DW-EIS, and dash-dotted~(brown) curves for
FBA.}\label{Fig04}
\end{center}
\end{figure*}
 In most of the case studies of the typical ion-atom
collisional reactions, it has been shown that the inter-nuclear
interaction plays a major role in the differential cross sections,
while the role changes to a minor contribution, or not influence at
all, in the integrated cross sections. Here, at least for most of
the present considered cases, the comparisons show that the role of
the inter-nuclear interaction is not crucial even in the fully
differential cross sections.\par
Interestingly, the theory predicts a double peak structure for
($E_e=10$~eV, $q=1~a.u.$) around $\phi_e=180^\circ$. The second peak
predicted by 3DW-EIS is higher than that is predicted by CDW-EIS.
Although double-peak structure seems to be real, it is not seen
vividly in the measured cross sections probably for the low
resolution of the experiments. However, to further confirm the
reality of this theoretically predicted behavior further experiments
with better resolution should be carried out. It should be noted
even if the predicted structure be real, the binary peak in the left
part of the panel is slightly more shifted to left than experimental
data. The present calculations confirm this hypothesis that the
predicted double-peak structure is attributed to the angular
distribution of the initial bound-state wavefunction of the ejected
electron and the kinematics only. However, the general kinematical
conditions under which occurrence of such a structure is predictable
are still questionable. For ($E_e=10$~eV, $q=0.4~a.u.$) the peak
position shifts slightly to left with respect to the 3DW-EIS
predictions and the measured data.\par
 It may be asserted that the reasonable agreement of the
obtained results with experiment is not surprising, since even the
first Born approximation~(FBA) is expected to be reasonable at
sufficiently high impact energies. In order to examine the validity
of such a hypothesis, we compared the $2s$-ionization results for
($E_e$=10~eV, $q=0.4~a.u.$) and ($E_e$=20~eV, $q=0.4~a.u.$) with the
FBA approximation in figure~4. As is seen, at the angular regions
far from the binary peak, the agreement of FBA with experiments is
very poor, while the CDW-EIS and 3DW-EIS results are in reasonable
agreement with the measurements in these angular regions.\par
\begin{figure}[t]
\begin{center}
\includegraphics[scale=1]{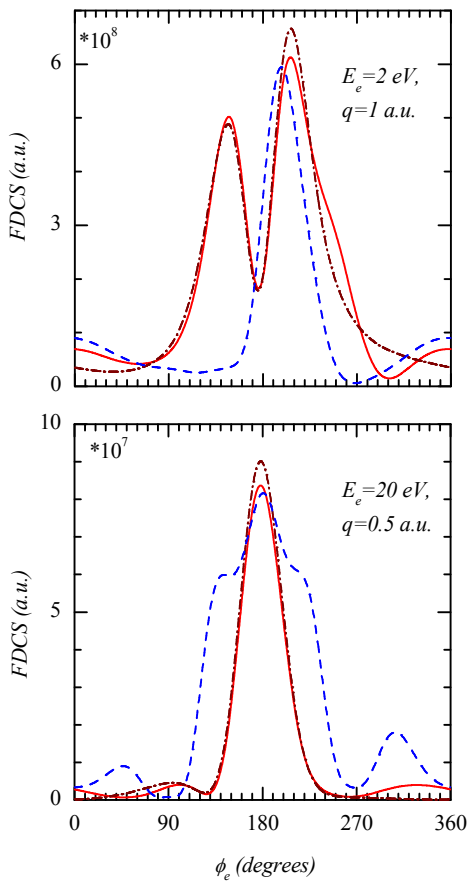}
\caption{(Color online) Comparison of the full post and prior
CDW-EIS calculations with the 3DW-EIS results. Solid~(red) curves
are for the prior CDW-EIS, dash-dotted~(brown) curves for the prior
CDW-EIS and dashed~(blue) curves for 3DW-EIS.}\label{Fig05}
\end{center}
\end{figure}

Comparison of the theoretical predictions for ionization of $Li(2p)$
in the cases ($E_e=2$~eV, $q=1~a.u.$) and ($E_e=20$~eV,
$q=0.5~a.u.$) is instructive, although there do not exist available
experimental cross sections for these cases. This comparison is
performed in figure~\ref{Fig05}. The full post and prior CDW-EIS
cross sections and the 3DW-EIS results are depicted in this figure.
For both cases, the NN interaction does not considerably change the
cross sections neither in shape nor in magnitude, while the
post-prior discrepancy is complectly obvious in some angular
regions. However most of the aspects in the post and prior graphs
are the same.\par
For these cases, the 3DW-EIS~\cite{Ghanbari2} and CDW-EIS theories
predict totally different behaviors for the ejected electron
spectra.  Such an obvious difference has been also reported between
3DW-EIS and FBA~\cite{Ghanbari2}.  For ($E_e=2$~eV, $q=1~a.u.$),
CDW-EIS predicts a double binary peak structure, while the 3DW-EIS
predicts a single right-shifted binary peak. Also, the width of the
peak structures is very different in both predictions. Such a double
peak structure is predicted by both theories for ($E_e=10$~eV,
$q=1~a.u.$). According to the 3DW-EIS predictions for that case, the
lower peak goes from left to right passing through the classically
expected peak position when the projectile momentum transfer is kept
constant at $q=1~a.u.$ but $E_e$ increases gradually from 4~eV to
16~eV~\cite{Ghanbari2}. We examined this point using the CDW-EIS
theory, and found that the double peak structure appears even for
emission energies lower than 1~eV. Interestingly, such a similar
situation was observed when we kept $q$ at 1.2~$a.u.$ and changed
$E_e$ in the same range. Also, for higher values of $q$, such a
double peak structure is observable at various energy intervals. For
example, two peaks with nearly the same heights is observed for
$q=1.2,~1.5$ and 2~$a.u.$ at $E_e=8,~18$ and 40~eV, respectively.
This means that in the CDW-EIS and 3DW-EIS theories the kinematical
conditions under which the double peak structure appears are
different. So, further experimental studies are necessary to
identify the correct conditions.\par
For ($E_e=20$~eV, $q=0.5~a.u.$), this double-peak structure
disappears, instead, two nearly mirror-symmetric shoulders appears
in both sides of the binary peak in the 3DW-EIS cross sections. In
the post and prior CDW-EIS results these shoulders are of course
absent and only a relatively sharp peak is seen. Also, both theories
predict some structures in angular domains of $[0^\circ,90^\circ]$
and $[270^\circ,360^\circ]$ in both side of the peak region.
Post-prior discrepancy is relatively large in these angular regions
and also at the binary peak angle. Also for this case, the results
of the 3DW-EIS and CDW-EIS theories are obviously different and
since there is no experimental data in hand, we cannot conclude
which one of the theoretical methods can explain the mechanism in
these cases better.\par
It may be discussed that since the 3DW-EIS and CDW-EIS do not employ
the same electronic states, it is difficult to judge whether the
extra term of the quantum transition amplitude or the different
approximations for the electronic states is the main source of the
differences observed in the calculations in some cases. In order to
clarify this point, we recall that the 3DW-EIS results are not so
sensitive to the radial part of the wavefunctions~\cite{Ghanbari2}.
Since the angular part of the employed wavefunctions are the same in
both theories, so the difference of the electronic wavefunctions
cannot be the source of the discrepancies observed between the
3DW-EIS and CDW-EIS results. However, it is worthy to point out
that, in some cases, using the same numerical HF~wavefunctions
employed in 3DW-EIS, the CDW-EIS theory produces the results in very
good accordance with the quantum
considerations~\cite{Voitkiv2,Voitkiv3,Voitkiv4}. For example, a
semiclassical CDW-EIS approach with numerical
HF~wavefunctions~\cite{Voitkiv4} is recently used to simulate a
recent precise experiment~\cite{Gassert} and a very good agreement
with the experiment was found.\par
 It may be argued that with the study based on the fully
quantum-mechanical version of CDW-EIS already reported in
Ref.~\cite{Ghanbari2}, there is no room for publication of the
present study with two more approximations: 1)~classical
straight-line motion for the projectile, 2)~the employment of the
hydrogenic Coulomb wave functions for the description of the target
of $Li$. In order to rebut such a presumption, we would like to
highlight several points as follow: i)~As the reader affirms, the
present model is a semi-classical model while the other one is a
fully quantum-mechanical approach. So the calculations are very
different but in most of the cases the results are close to each
other. ii)~In this contribution, we focused on the NN~interaction
and for some specified cases it has been shown that  the
NN~interaction plays a minor role in the collision dynamics even the
fully differential cross section is the case.  iii)~In this paper,
we investigated the post-prior discrepancies observed in the
calculated fully differential cross sections, while such a
discussion is absent in Ref.~\cite{Ghanbari2}. We showed for some
cases this discrepancy is much more important than the
nuclear-nuclear interaction. iv) Here, we showed that for some
cases, FBA is not a good theory, although the impact energy is high
enough. v) In the present work, it has been shown that for some
values of $E_e$ and~$q$, the predictions made by the 3DW-EIS and
CDW-EIS are very different, and in Ref.~\cite{Ghanbari2} it has been
shown that for the same cases the 3DW-EIS results are clearly
different from~FBA results. Consequently, there are some cases where
3DW-EIS, CDW-EIS and FBA predictions are very different. vi) Both
3DW-EIS and CDW-EIS theories predict a double binary peak structure
for some cases. However, we showed that the kinematical conditions
leading to such a structure are different in the theories. For the
last two pints, further experimental data is needed to judge the
validity of the theoretical predictions.
\section{Conclusions \label{Sec04}}
The three-body CDW-EIS method was used to theoretical study of the
single ionization of the neutral lithium atoms in the $Li^{2+}-Li$
collision systems. Detachment of the outer electron from $Li(2s)$
and $Li(2p)$ by impact of 16~MeV $Li^{2+}$ ions was considered. The
FDCSs for ejection of the outgoing electron into the azimuthal plane
were calculated as a function of the ejected electron's azimuthal
angle. Both post and prior cross sections were evaluated. For
different values of the ejected electron energy~($E_e$) and the
projectile momentum transfer~($q$), the obtained results were
compared with the similar calculations performed using the quantum
version of the formalism. This later version of the theory is
labeled as 3DW-EIS in the literature to distinguish from CDW-EIS.
Also, in the cases for which experimental data is available the
results were compared to those data.\par
For ionization of $Li(2s)$, in the cases for which the binary peak
is relatively sharp and it is well demystified in the measurements,
the agreement of the results with the 3DW-EIS and with experiments
is very good. For such cases, the NN interaction does not play a
significant role in the break up process. Also, the post-prior
discrepancy is small for these cases. But, for the cases that the
experimental data are too diverse to clearly illustrate the binary
peak, some discrepancies was found between CDW-EIS and 3DW-EIS
theories. For these cases, both the role of the NN interaction in
the reaction and the influence of the post-prior discrepancy on the
cross sections become more obvious.\par
For ionization of $Li(2p)$, in the cases for which there exist
experimental data, the overall agreement between CDW-EIS and
experiment is reasonable. However, some small differences are
observed between CDW-EIS and 3DW-EIS. For example, for ($E_e=10$~eV,
$q=0.4~a.u.$), these differences were observed both in the shape and
in the peak position predicted for the angular distribution of the
cross sections by these theories. Also, for ($E_e=10$~eV,
$q=1~a.u.$), both theories predict a double-peak structure for the
cross sections, but the second peak predicted by 3DW-EIS is much
higher than that of CDW-EIS. Our calculations showed that this
structure is attributed to the angular part of the initial bound
state wavefunction of the active electron and to the kinematical
conditions. In contrast to these differences the overall agreement
of the theories seems fair. For, these cases both the NN interaction
and the post-prior discrepancy play a minor role in the calculated
results.\par
The major difference between the CDW-EIS and 3DW-EIS theories takes
place for ionization of $Li(2p)$ in the cases for which there is no
experiential data. For ($E_e=2$~eV, $q=1~a.u.$), CDW-EIS predicts a
double-peak structure, while such a complex structure is absent in
the 3DW-EIS predictions. Also, 3DW-EIS predicts two shoulders in
both sides of the binary peak, while these shoulders are absent in
the CDW-EIS results. Consequently the kinematical conditions for
occurring a double-peak structure are different in these theories.
Further experiments are needed to indicate which one of these sets
of predictions are real. For these cases, the NN interaction is not
important but the post-prior discrepancy is considerable in some
angular regions and at the binary-peak angle.
\begin{acknowledgments}
OG would like to acknowledge the office of graduate studies at the
University of Isfahan for their support and research facilities.
\end{acknowledgments}

\end{document}